\documentclass[usenatbib,usegraphicx,onecolumn]{mn2e}
\usepackage[english]{babel} 
\usepackage{color}
\usepackage{ulem} 

\def\mnras{MNRAS}

\def\aap{A \& A}

\def\V2{V_2}
\def\V2ij{V_{2ij}}
\def\S{{\mathcal S}}
\def\V{\mathcal{V}}

\def\la{\left\langle}

\def\lsim{~\rlap{$<$}{\lower 1.0ex\hbox{$\sim$}}}

\def\gsim{~\rlap{$>$}{\lower 1.0ex\hbox{$\sim$}}}

\usepackage{graphics}
\usepackage{color}
\usepackage{amsmath}
\usepackage{amssymb}
\usepackage{psfrag}
\usepackage{subfig}
\usepackage{graphicx}

\newsavebox{\measurebox}
\begin{document}
\date {} 
\title[Supernova remnant power spectrum] {Angular power spectrum of supernova remnants: effects of structure, geometry and diffuse foreground}
\author[Choudhuri et al.]{
Samir Choudhuri,$^{1}$\thanks{E-mail: s.choudhuri@qmul.ac.uk}
Preetha Saha,$^{2}$
Nirupam Roy,$^{3}$
Somnath Bharadwaj,$^{2}$
and
\newauthor Jyotirmoy Dey $^{4}$
\\
$^{1}$Astronomy Unit, Queen Mary University of London, Mile End Road, London E1 4NS, United Kingdom \\
$^{2}$ Department of Physics and Centre for Theoretical Studies, Indian Institute of Technology, Kharagpur 721302, India\\
$^{3}$ Department of Physics, Indian Institute of Science, Bangalore 560012, India\\
$^{4}$Department of Earth and Space Sciences,Indian Institute of Space Science and Technology, Trivandrum 695547, India
}
\maketitle

\begin{abstract}
The study of the intensity fluctuation power spectrum of individual supernova remnants (SNRs) can reveal the structures present at sub-pc scales, and also constrain the physical process that generates those structures. There are various effects, such as the remnant shell thickness, projection of a three-dimensional structure onto a two-dimensional observational plane, and the presence of diffuse ``foreground'' emission, which causes the observed power spectrum to deviate from the intrinsic power spectrum of the fluctuations. Here, we report results from a systematic study of these effects, using direct numerical simulations, in the measured power spectrum. For an input power-law power spectrum, independent of the power-law index, we see a break in the observed power law at a scale which depends on the shell thickness of a shell-type SNR, and the three-dimensional turbulence changes to two-dimensional turbulence beyond that scale. We also report how the estimated power spectrum is expected to deviate from the intrinsic SNR power spectrum in the presence of additional diffuse Galactic synchrotron emission (DGSE) around the remnant shell. For a reasonable choice of the parameters, if the intrinsic SNR power spectrum is shallower than the DGSE power spectrum, the SNR contribution dominates at small angular scales of the estimated power spectra. On the other hand, if the SNR power spectrum is relatively steeper, the original power spectra is recovered only over a small window of angular scales. This study shows how detailed modeling may be used to infer the true power spectrum from the observed SNR intensity fluctuations power spectrum, which in turn can be used to constrain the nature of the turbulence that gives rise to these small scale structures.
\end{abstract} 


\section{Introduction}
\label{intro}

Supernova remnants (SNR) are the structures generated by the explosion at the end stage of a massive star. In this process, a tremendous amount of mass and energy is injected in the interstellar medium (ISM), and it causes the heating of the ISM. The ISM also gets enriched by heavy elements created in the stellar core or in the supernova explosion. The SNRs also interact with the surrounding molecular clouds \citep{uchiyama02,slane15,sano18,sano20}. There are more than 295 known SNRs in the Milky Way Galaxy \citep{green14}. The SNRs are one of the key factors to shape the large scale structure and the properties of the ISM \citep{deavillez04,hill12,hennebelle14,gatto15}. The overall morphological study of the SNR can reflect the characteristics of the progenitor supernova explosion \citep{lopez11,orlando16}. The detailed interaction of the SN blasts with the inhomogeneous ambient medium can be studied using the morphology \citep[e.g.]{zhang19}, as well as the small scale structures of SNRs. For example, \citet{stafford19} found that larger (older) SNRs are more elliptical/elongated as SNR shells become more asymmetric as they sweep up the ISM. On the other hand, the small scale structures of the SNR are expected to reveal the nature of the turbulence in the remnant plasma \citep{roy09,saha19}. 

Numerical simulations of supernova explosion and evolution of the SNRs are useful to predict the observable properties of the SNR, e.g., morphology, the abundance of the heavy elements in the ISM, properties of the progenitor and the compact remnant object, etc. The three-dimensional hydrodynamic simulation of the SNR has been used to predict the SN light curve \citep[e.g.][]{ultrobin17,ultrobin19}, as well as to estimate the kinetic energy of the explosion, type of the progenitor, and the ambient medium density \citep[e.g.][]{badenes08,patnaude15,yamaguchi14}. \citet{orlando16} simulate SNR similar to Cassiopeia~A (Cas~A), to derive the energies and masses of the post-explosion anisotropies. Recently \citet{zhang19} hydro-dynamically simulate the interaction between an individual supernova remnant (SNR) and a turbulent molecular cloud medium and found that the properties of SNRs are mainly controlled by the mean density of the surrounding turbulent medium. A comparison of the observed small scale structures with that from the numerical simulations may be useful to understand the evolution of SNRs and their interaction with the surrounding medium. However, this requires detailed quantification of the SNRs' small scale structures. 

Although there are many reports in the literature regarding the large scale structure of the Galactic SNR, only a few exists which study the fine-scale structure of the SNR. \citet{roy09} used the radio continuum of Cas~A and Crab SNRs, and studied the synchrotron emission fluctuations at a wide range of angular scales. The synchrotron emission is produced by the relativistic electrons in the magnetic field \citep{berezhko04}, and the scale-free fluctuations are likely to be produced by the magnetohydrodynamic turbulence in the synchrotron emitting plasma. They also observed a break in the measured power spectrum for the shell-type SNR Cas~A due to transition from three- to two-dimensional turbulence at scale corresponds to the shell thickness of Cas~A. Recently, \citet{saha19} have extended this work and studied the power spectrum of the Kepler SNR. They also have found a break due to the shell thickness of the Kepler SNR. Interestingly, they have found another break in the power spectrum at even smaller scales less than the shell thickness. As concluded in \citet{saha19}, this might be due to the dominant contribution from the Galactic or extra-galactic foregrounds.

The inferred power spectrum of the intensity fluctuations from observations of SNRs (radio synchrotron emission, but also at other wavelengths) might change due to various effects. As shown in \citet{roy09} and \citet{saha19}, the intrinsic slope of the SNR power spectrum is changed due to the shell thickness of a shell-type SNR. The projection of a three-dimensional structure in a two-dimensional observational plane can modify the shape of the intrinsic power spectrum generated by the actual magnetohydrodynamic turbulence. Another physical process that might change the intrinsic power spectrum is the presence of stray radiation from the ambient medium (e.g., Galactic synchrotron emission). Due to the presence of these effects, it is challenging to probe the intrinsic SNR power spectrum. 

In this work, we have systematically analyzed the effects of overall SNR morphology (with different shell thickness), projection, and presence of stray emission to quantify how all these change the inferred power spectrum for a given input power spectrum. Here, we follow a forward modeling approach to accommodate each of these effects in our simulation, and to understand how the intrinsic power spectrum can be constrained from real observations. We start with the toy model of the SNR with fluctuations modeled as a three-dimensional Gaussian random variable (GRF) and a known input power-law power spectrum. We note that, although there are several hydrodynamical simulation approaches to model the small scale structures in SNRs (e.g., \citealt{luo20} for Crab nebula), this simple GRF model for simulation is sufficient for the purpose of the current work to study the aforementioned effects. For different input power-law power spectra with a range of power-law index, we systematically study the effect of the shell geometry and the projection of the three-dimensional shell-type structure in a two-dimensional image plane, on the final estimated power spectra. Finally, we have also studied how the presence of the diffuse Galactic synchrotron emission affects the SNR power spectrum measurements.

Please note that this work is part of a comprehensive study of quantifying and understanding the small scale structures present in SNRs. This work presents the details of a systematic forward modeling approach that can be adopted to infer the intrinsic power spectra from the measured power spectra. An application of this method for the specific case of the radio synchrotron emission from Kepler SNR has been reported in \citet{saha19}. Further implementation for radio and X-ray power spectra, as well as cross power spectra, will be presented in future work (Saha et al., 2020 (submitted)). A brief outline of the current paper is as follows. In Section 2, we describe the simulation used and also the power spectrum estimation algorithm. In Section 3, we describe the results for the different geometric effects of the SNR. In Sections 4, we study the impact of the Galactic synchrotron emission in the measured SNR power spectrum. Finally, we present discussion and conclusions in Section 5.

\section{Simulation}
\label{sim}
\begin{figure*}
\begin{center}
\includegraphics[width=60mm,angle=0]{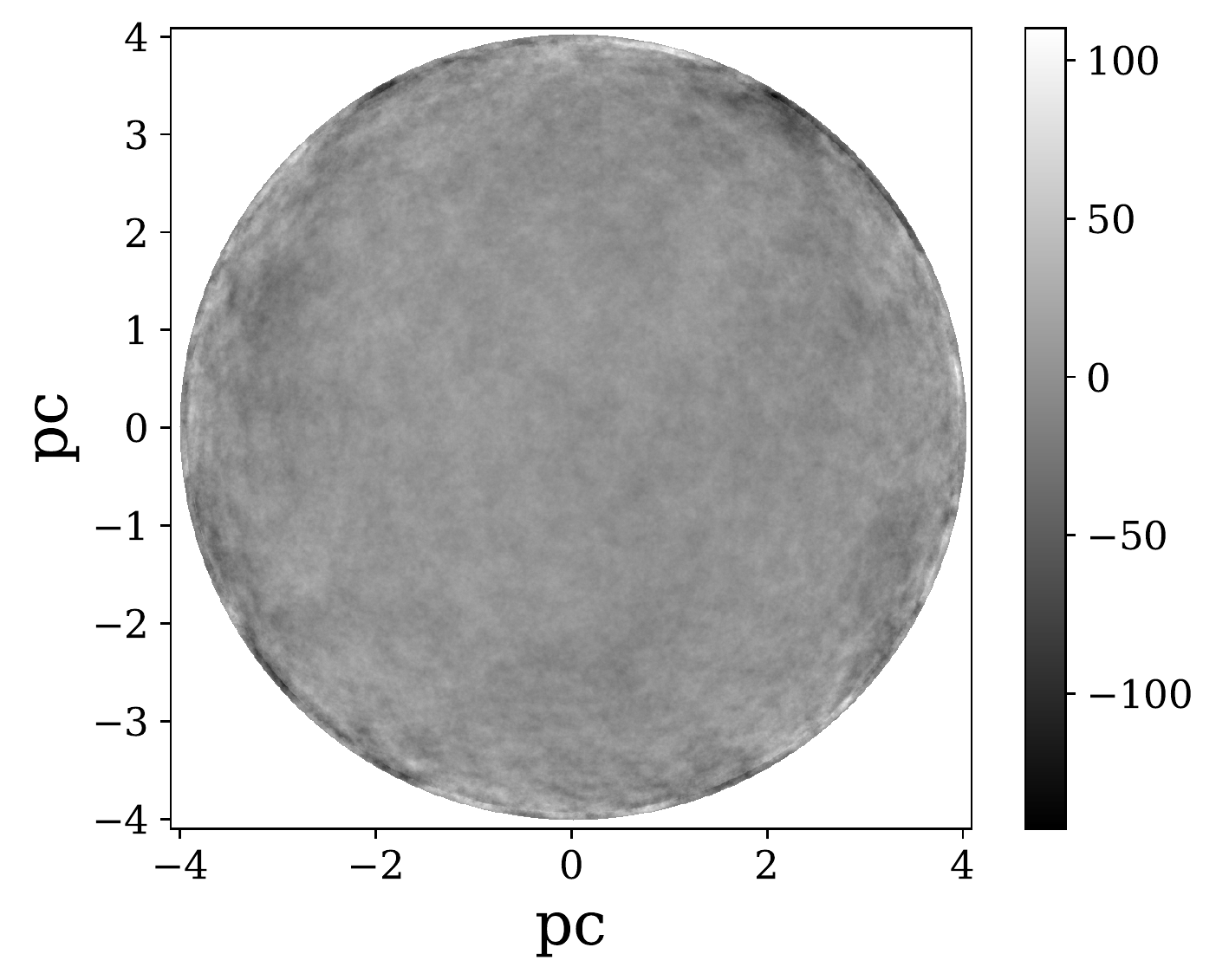}
\includegraphics[width=55mm,angle=0]{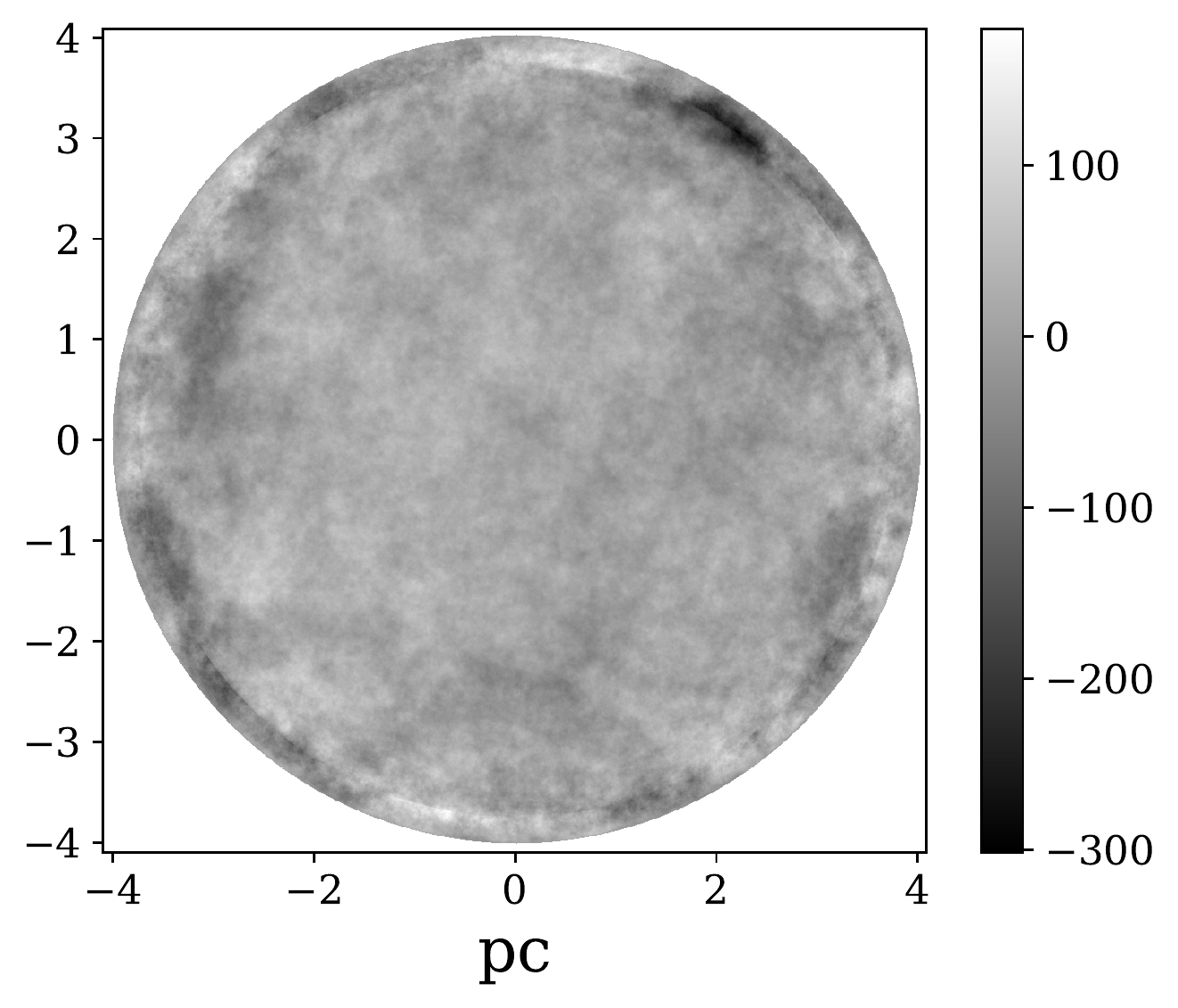}
\includegraphics[width=55mm,angle=0]{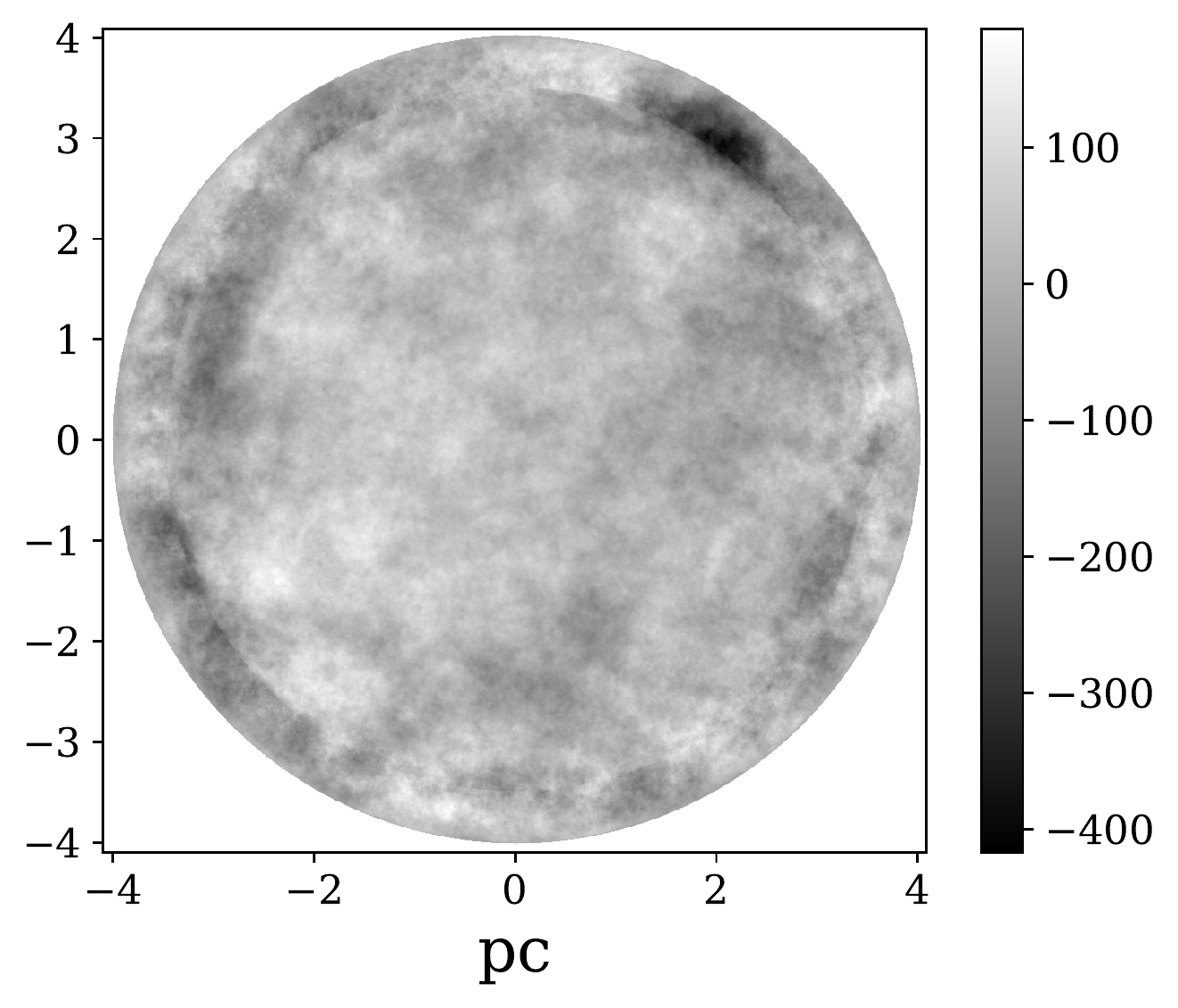}
\caption{The simulated shell type SNR with different shell thickness. The left, middle and right panels are for shell thickness 8, 32 and 64 pixels respectively. Here, we use $\beta=-3$ and we collapse all the slices along the line of sight (LoS). Here the color bars represent the surface brightness of the SNRs but in arbitrary unit.}
\label{fig:fig1}
\end{center}
\end{figure*}

In this section, we describe the simulation process used to generate the SNR image. We assume the emission fluctuations of the SNR as a 3-d Gaussian random field (GRF) with a power-law power spectrum. To simulate, we follow the same procedure, as discussed in \citep[Sec.~4,][]{samir16b}. First, we generate the fluctuations in a three-dimensional grid in the Fourier domain using
\begin{equation}
\Delta ({\bf k})=\sqrt{\frac{ V \, P(k)}{2}}[a({\bf k})+ib({\bf k})],
\label{eq:1}
\end{equation}
 where $V$ is the volume of the three-dimensional cube, $\Delta ({\bf k})$ is the fluctuations in the Fourier domain. Here, $P(k)=k^{\beta}$ is the power spectrum at wavenumber $k$, and $a(\bf{k})$ and $b(\bf{k})$ are independent Gaussian random variables with zero mean and unit variance. We use the Fastest Fourier Transform in the West (FFTW; \citealt{frigo05}) to generate the fluctuations $\delta(x)$ in the image plane from $\Delta (\bf{k})$. Total number of pixel used here is $1024^3$ (corresponding to a box size of $(8.2~\rm{pc})^3$ with $8\times10^{-3}$ pc resolution). The parameters used here represent the typical size of an observed SNR (e.g., Cas~A). We multiply the simulated three-dimensional cube with an appropriate window to make structures for filled-center or shell-type SNR (with different shell thickness). For example, to generate a shell-type SNR, we choose a window that has a value unity within the shell and zeroes elsewhere. Figure \ref{fig:fig1} shows images of simulated realization of SNRs with different shell thickness. The left, middle and right panels of Figure \ref{fig:fig1} are for shell thickness 8, 32, and 64 pixels respectively. In this figure, we use an input power-law index $\beta=-3$ and collapse all the slices along the line of sight (LoS) to create the projected two-dimensional image.
 
We use the FFTW to convert the line of sight collapsed two-dimensional image to the Fourier plane, and then square each Fourier mode to estimate the power spectrum. For smaller field of view, the angular power spectrum is same as the power spectrum, and we are using the notation $P(k)$ throughout this paper.  We divide the whole k-range $(k= 2\pi/ L)$ into 20 equally spaced logarithmic bins to increase the signal to noise ratio. Here we use the k range 1.08 to $\sim 500 {\rm pc^{-1}}$ in this analysis. The results for the power spectrum analysis are given in the next section.

\section{Results}
\label{result}
\begin{figure*}
\begin{center}
\hspace*{-1.25cm}
\includegraphics[width=200mm,angle=0]{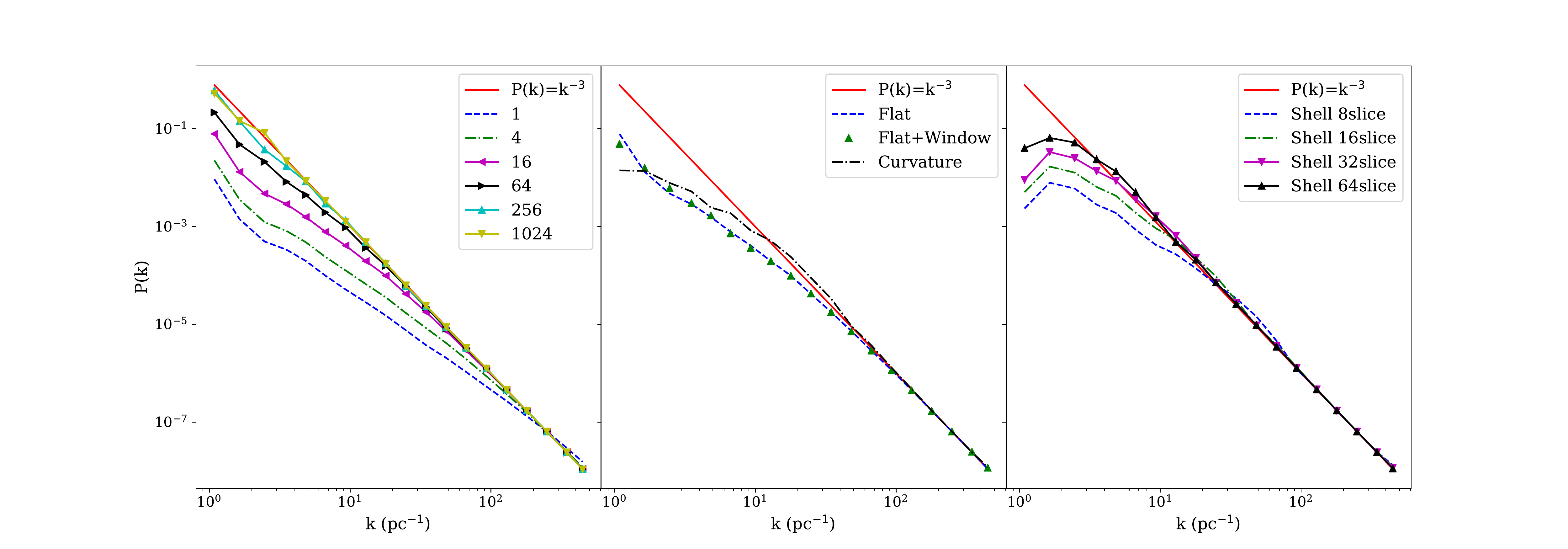}
\caption{{\it Left}: The measured power spectra for different number of slices (marked with different color) collapsed together along the line of sight. Here input power spectrum is $P(k)=k^{-3}$ (red line) for these simulations. The break in the power spectra shifts towards small $k$ as the number of slices is increased. {\it Middle}: The effect of curvature in the measured power spectrum. The blue dashed line (green triangles) is power spectrum derived from image created by collapsing 16 slices along the LoS without any (with a) circular window. The black dashed-dot line is for a shell-type structure with a shell width of 16 pixels collapsed along the LoS. {\it Right}: The combined effect of shell thickness and curvature in the measured power spectrum. The estimated power spectra corresponding to shell thickness of 8, 16, 32 and 64 pixels are shown with different color}
\label{fig:fig2}
\end{center}
\end{figure*}

As a first step, we study the effect of collapsing multiple slices along any given line of sight. For this, we use the full box without the SNR window. The estimated power spectra after collapsing different number of slices ($1,~4,~16,~64,~256,~1024$) are shown in Figure \ref{fig:fig2} (left panel). The solid red line shows the input power spectrum $P(k)=k^{-3}$ used for the simulation. We normalize the amplitude of the measured power spectrum at the largest $k$ mode for clear visualization. As expected, for a very narrow line of sight extent (i.e., fewer slices), the measured power-law index deviates from the input index. The reason for this is that effectively we are measuring the two-dimensional fluctuations where the power-law index is expected to be shallower by 1. As the thickness along LoS being averaged increases, we measure 3-d fluctuation power spectrum at scales smaller than the thickness, but 2-d at larger scales (smaller $k$ range). In the left panel of Figure \ref{fig:fig2}, we clearly see a break in the power-law due to the transition from the three-dimensional fluctuations to two-dimensional at larger scales, and the break shifts towards the left when the number slices collapsed along LoS is increased. Only for $k > k_{\rm break}$, we are probing the inherent three-dimensional fluctuations.

Next, we study the effect of curvature in the measured fluctuations power spectrum of the collapsed 2-d SNR image. For this, we simulate a hemispherical shell with fixed radial shell thickness (16 pixels in the example shown), and collapse that along the LoS to a 2-d image (``Case I''). Note that, due to the curvature, this collapsing results in a systematic variation of LoS thickness at a different radius of the 2-d image. The projection also changes the observed 2-d angular separation between two points from the true one in a non-trivial way depending on the positions and orientation; hence the inferred 2-d power spectrum is expected to be altered. For comparison, we take a second image, as earlier by collapsing the same number of slices (16 in this case), but multiplying it with a circular window whose diameter is the same as the outer radius of the spherical shell (``Case II''). The results are shown in the middle panel of Figure \ref{fig:fig2}. The input power law and the estimated power law for Case I and II are shown in red, black lines and green triangles respectively. The measured power spectra without curvature effect are quite similar without (blue line) or with the circular window (green triangles), indicating that the effect of the window is negligible in the measured power spectrum. Basically, the power spectrum will be convolved with the Fourier transform of the window \citep{samir14}. Here, the width of the window is very small in the Fourier domain; hence the effect is negligible. However, the curvature moves the break in the spectrum to smaller $k$ almost by a factor of $3$; we can recover the three-dimension input power spectrum in Case I for a large range $15<k<500~{\rm pc}^{-1}$ (compared to $50<k<500~{\rm pc}^{-1}$ in Case II). Also, the power spectrum at small $k$ is shallower compared to the earlier one. The right panel of Figure \ref{fig:fig2} shows the combined effect of both the variation of the shell thickness and the projection of a curved shell on a two-dimensional image on the measured power spectrum (for a shell thickness of 8, 16, 32, and 64 pixels). As expected, the break in the power spectra shifts towards smaller $k$ as the shell thickness is increased. For example, the black lower triangle for shell thickness 64 pixels can recover the input power spectrum for $k > 3 {\rm pc}^{-1}$, whereas the blue dashed line for 8 pixels shell thickness can recover for $k > 25 {\rm pc}^{-1}$. Note that in this realistic case, $k_{\rm break}$ is related to (but not exactly correspond to) the inverse of the shell thickness. 

\begin{figure*}
\begin{center}
\includegraphics[width=80mm,angle=0]{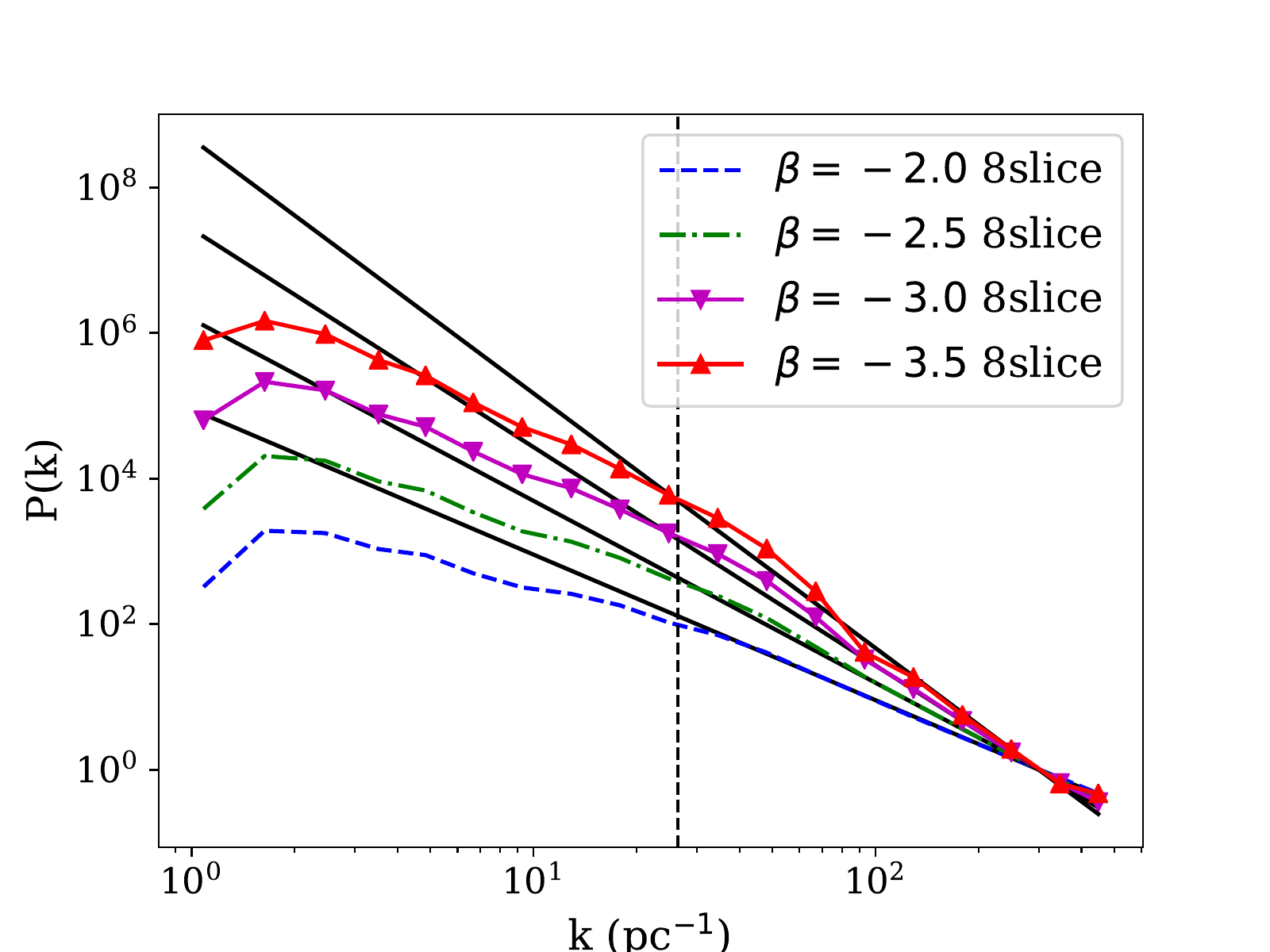}
\includegraphics[width=80mm,angle=0]{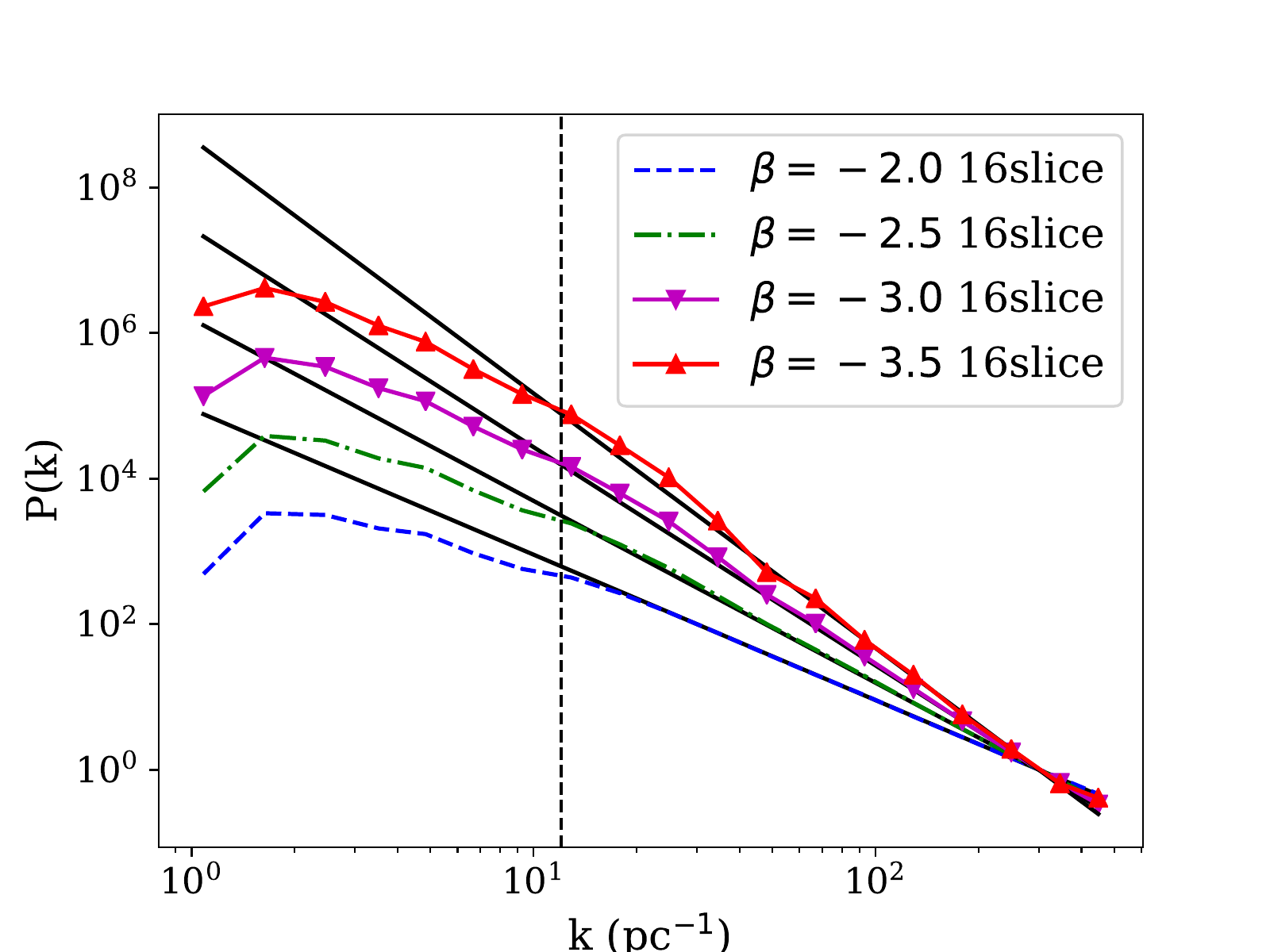}
\includegraphics[width=80mm,angle=0]{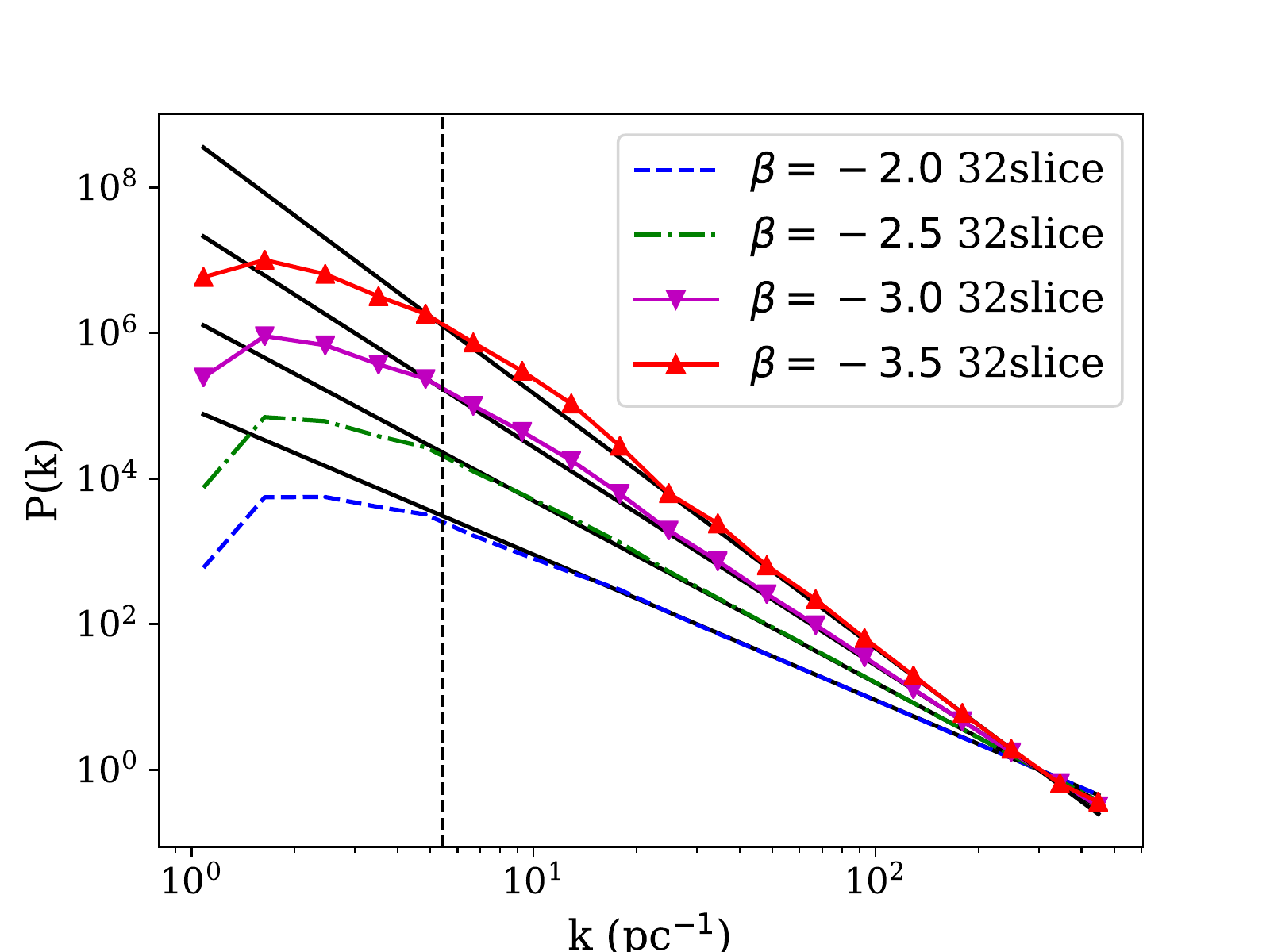}
\includegraphics[width=80mm,angle=0]{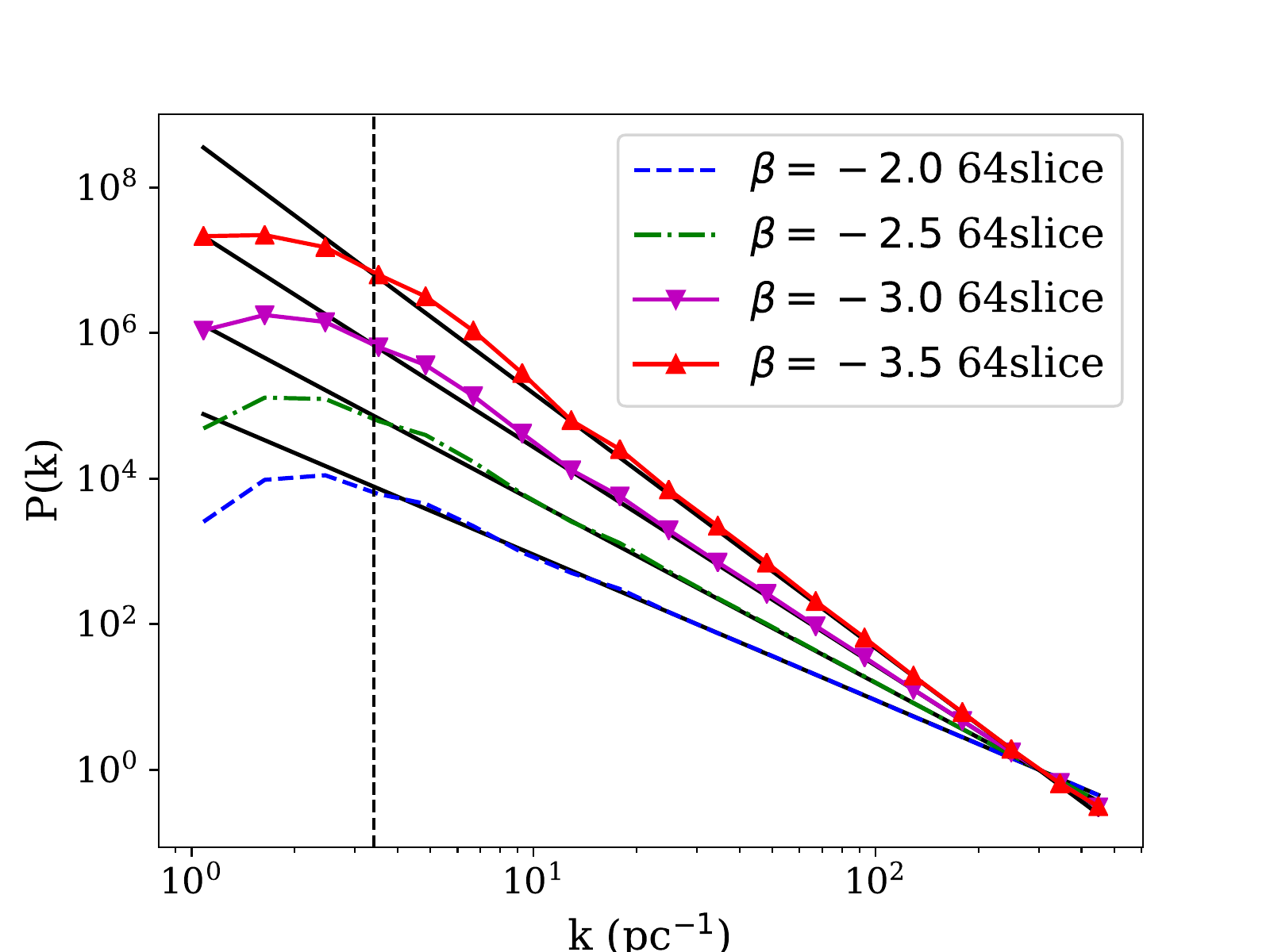}
\caption{The measured power spectrum for different shell thickness: 8 (top left), 16 (top right), 32 (bottom left) and 64 pixels (bottom right). The red, magenta, green and blue lines are for $\beta$ values -2, -2.5, -3 and -3.5 respectively. The solid black lines in each panel show the model power spectra with power law index $\beta$. The vertical black dashed lines in each panels show $k_{break}$ where the measured P(k) deviates from the input model. The value of $k_{\rm break}$ does not change with the input spectral index, but solely depends on the shell thickness.}
\label{fig:fig4}
\end{center}
\end{figure*}

Next, we vary the power-law index in the simulation and study how the measured power spectrum changes with shell thickness. We simulate for four values of $\beta=-2.0, -2.5, -3.0$ and $-3.5$. Here also we use four shell thickness of width $8, 16, 32$, and 64 pixels. Figure \ref{fig:fig4} summarizes the results. Here, each panel shows the estimated power spectrum for different shell thicknesses (8, 16, 32, 64 slices). Within each panel, we also show the results for different input $\beta$ values (shown as solid lines in each panel). We see that the estimated power spectrum deviates from the model at certain values of $k$ shown by the dotted vertical line in each panel. We also see that, for fixed shell thickness, the break remains at the same $k$ for all input $\beta$ values. For example, the value of the $k_{\rm break}$ is 26.4 ${\rm pc}^{-1}$ (top left panel) when the shell thickness is 8 pixels, and subsequently, it changes to 12.05 ${\rm pc}^{-1}$, 5.4 ${\rm pc}^{-1}$ and 3.4 ${\rm pc}^{-1}$ for 16, 32 and 64 pixels respectively. As explained earlier, when the shell thickness is more, we can measure the inherent three-dimensional fluctuations up to larger scales (smaller $k$). This behavior is quite similar for all values of $\beta$. Note, the value of $k_{\rm break}$ does not change with the input spectral index, but solely depends on the shell thickness.

\section{Effect of the stray radiation}
\label{dgse}
While estimating the angular power spectrum of a SNR, it is practically impossible to mitigate the contribution from diffuse stray radiation in the field. For example, at the low frequency, the diffuse Galactic synchrotron emission (DGSE), and its intensity fluctuations, will contribute to the detected continuum emission from the SNR. In this section, we study the effect of the DGSE in the measured SNR power spectrum, but this method can be generalized to other such stray emissions present in the field. For this, with the LoS collapsed image of the SNR, we add an additional diffuse component, present over the entire field of view. As the line of sight extent of the DGSE is much larger than that of the SNR, we add the 2D DGSE emission with the projected SNR image. There are four relevant parameters to model the SNR and the DGSE jointly: the (relative) amplitude of the SNR and DGSE power spectra, the intrinsic slope of the SNR power spectrum, the mean brightness of the SNR, and the mean brightness of the DGSE. The mean of the SNR can have some effects on the observed power spectrum at smaller $k$ values due to different thickness in the LoS collapsed image (discussed below). The mean component of the DGSE contributes only to k = 0 mode under the assumption of it being approximately constant over the small field of view, hence not included in the simulation. We fix the power spectrum of the DGSE to be a power-law with a power-law index of $2.34$ as constrained from low radio frequency observations \citep[e.g.][]{laporta08,bernardi09,ghosh150,samir17a,chakra19}. In these simulations, we use the SNR power spectra power law index of $\beta=$-1.5 and -3.0; this is to investigate both the scenario where the DGSE power spectrum is either steeper or shallower than that of the SNR.

\begin{figure*}
\begin{center}
\includegraphics[width=60mm,angle=0]{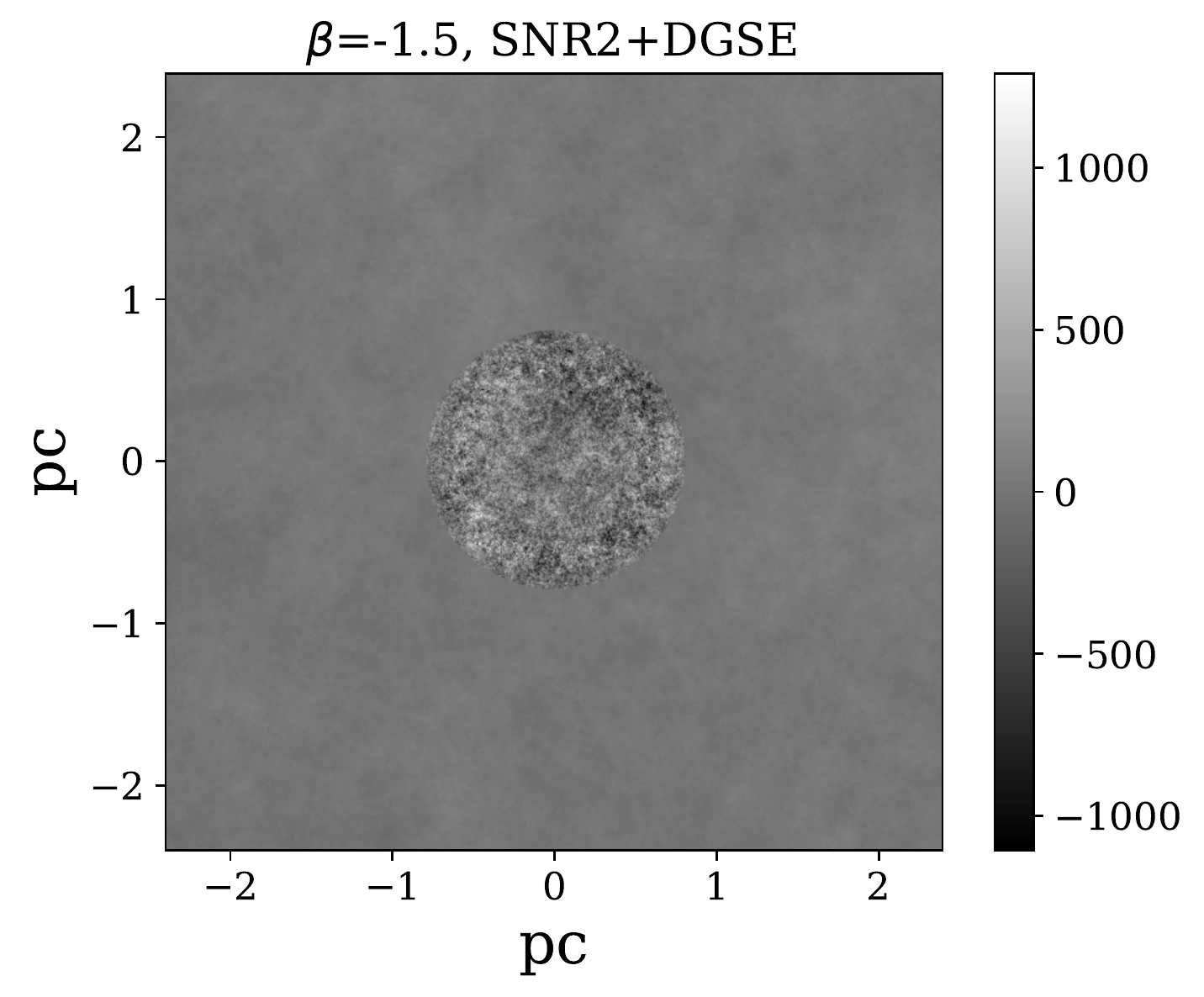}
\includegraphics[width=55mm,angle=0]{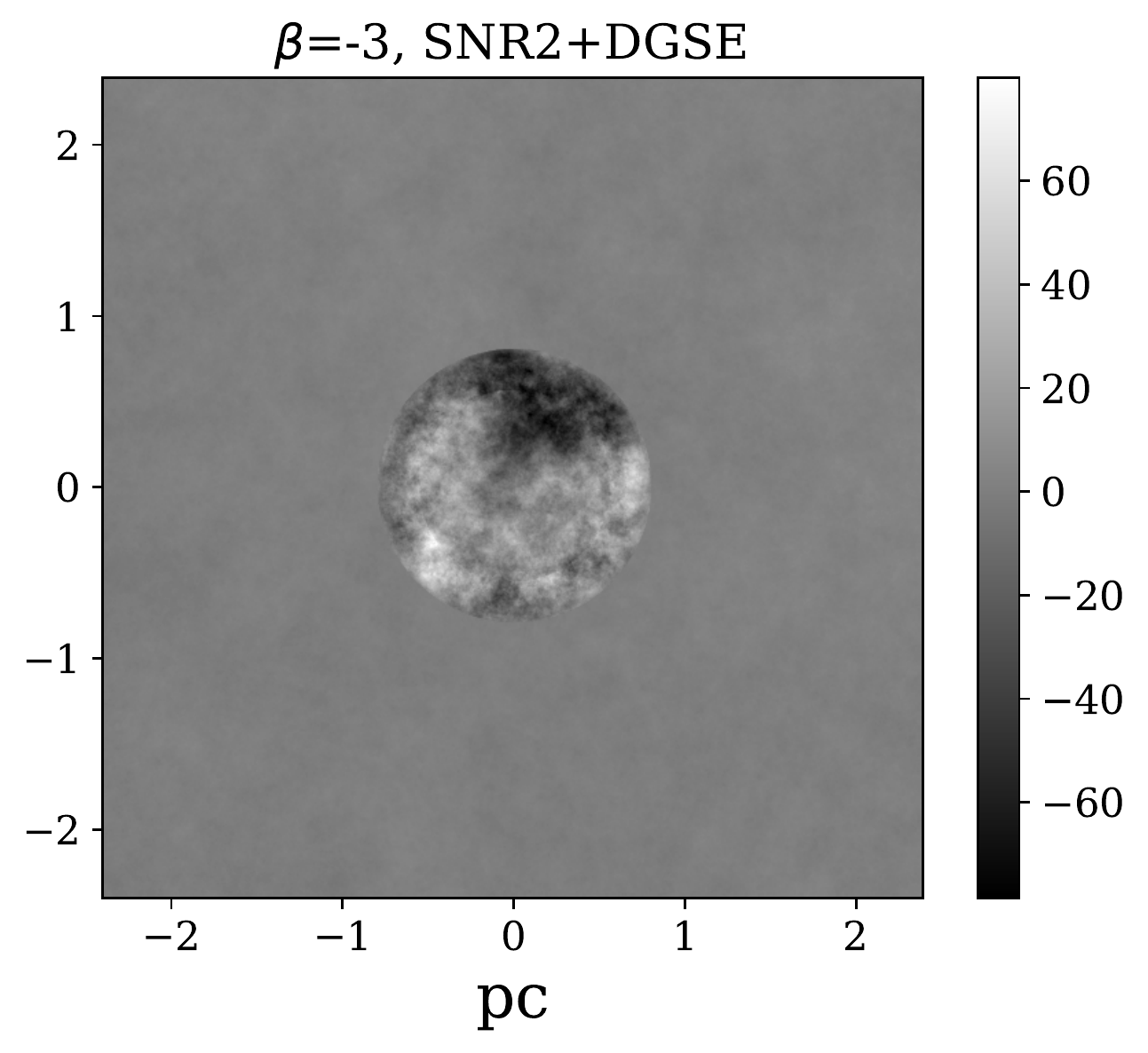}
\includegraphics[width=55mm,angle=0]{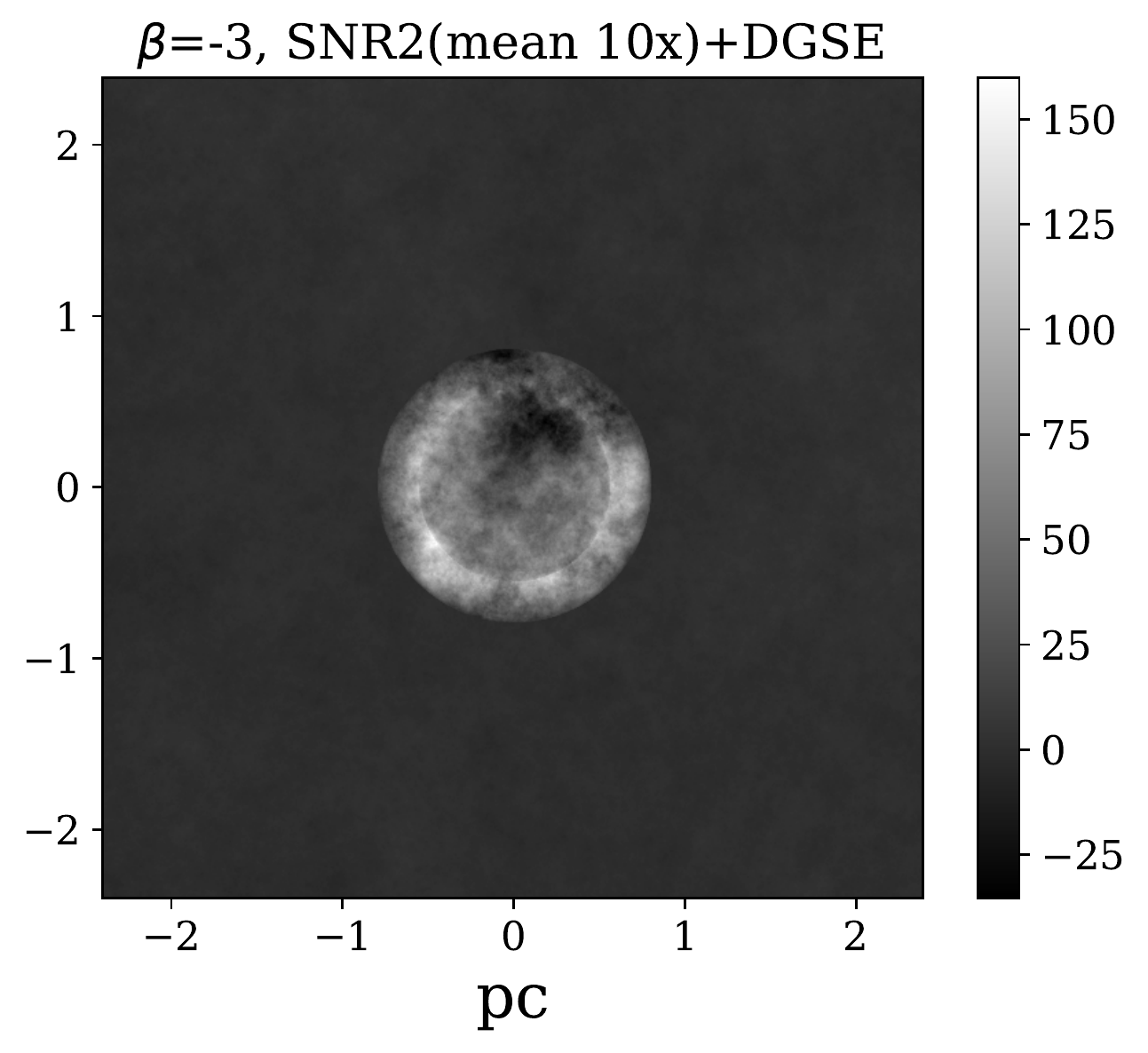}
\caption{Example realizations of the SNR in presence of the DGSE. The left and middle panels are for SNR power law index of -1.5 and -3.0, respectively. The right panel shows for -3 but the mean brightness of the SNR increased 10 times as compared with middle panel. The power law index of the DGSE is $-2.34$ for both the cases. The shell thickness of the SNR is 30 pixels (that corresponds to ``SNR2'' in Fig.~\ref{fig:fig5}). Here, the color bars represent the surface brightness of the SNRs but in arbitrary unit.}
\label{fig:fig5a}
\end{center}
\end{figure*}

To fix the dimensions of the simulation, we consider a realistic observation of the Kepler SNR at $\sim 1.4$ GHz (L-band) using the Giant Metrewave Radio Telescope (GMRT). The field of view (FoV) of the GMRT antenna at L-band is around $20^{'}$, and the diameter of the Kepler SNR is around $4^{'}$ \citep{patnaude12}. So, we choose the DGSE region with dimensions
$1024\times1024$ pixels and the diameter of the SNR for this simulation is 200 pixels (ratio $\sim 5$). We have used the parameters for the Kepler’s SNR (and GMRT) as a fiducial case, but compared the power spectrum for changes of various parameters (e.g. shell thickness to diameter ratio, input underlying 3-D Power Spectra, relative amplitude of DGSE and SNR power spectra etc.). The overall size of the SNR was kept unchanged, as that only introduces a scaling of the k axis, and does not change the overall nature of the power spectrum. Please note that the formalism and the suit of models can be tuned and scaled based on the actual value of these parameters estimated from the data (e.g. shell size, shell thickness, relative brightness etc.) while this method is used to model the underlying turbulent structure. Figure \ref{fig:fig5a} shows the LoS collased image of the SNR embedded within the DGSE. We vary the shell thickness and the power-law index of the SNR and the relative amplitude of the input DGSE and SNR power spectra to check the effect of DGSE on the estimated power spectra in different conditions. 

\begin{figure*}
\begin{center}
\includegraphics[width=80mm,angle=0]{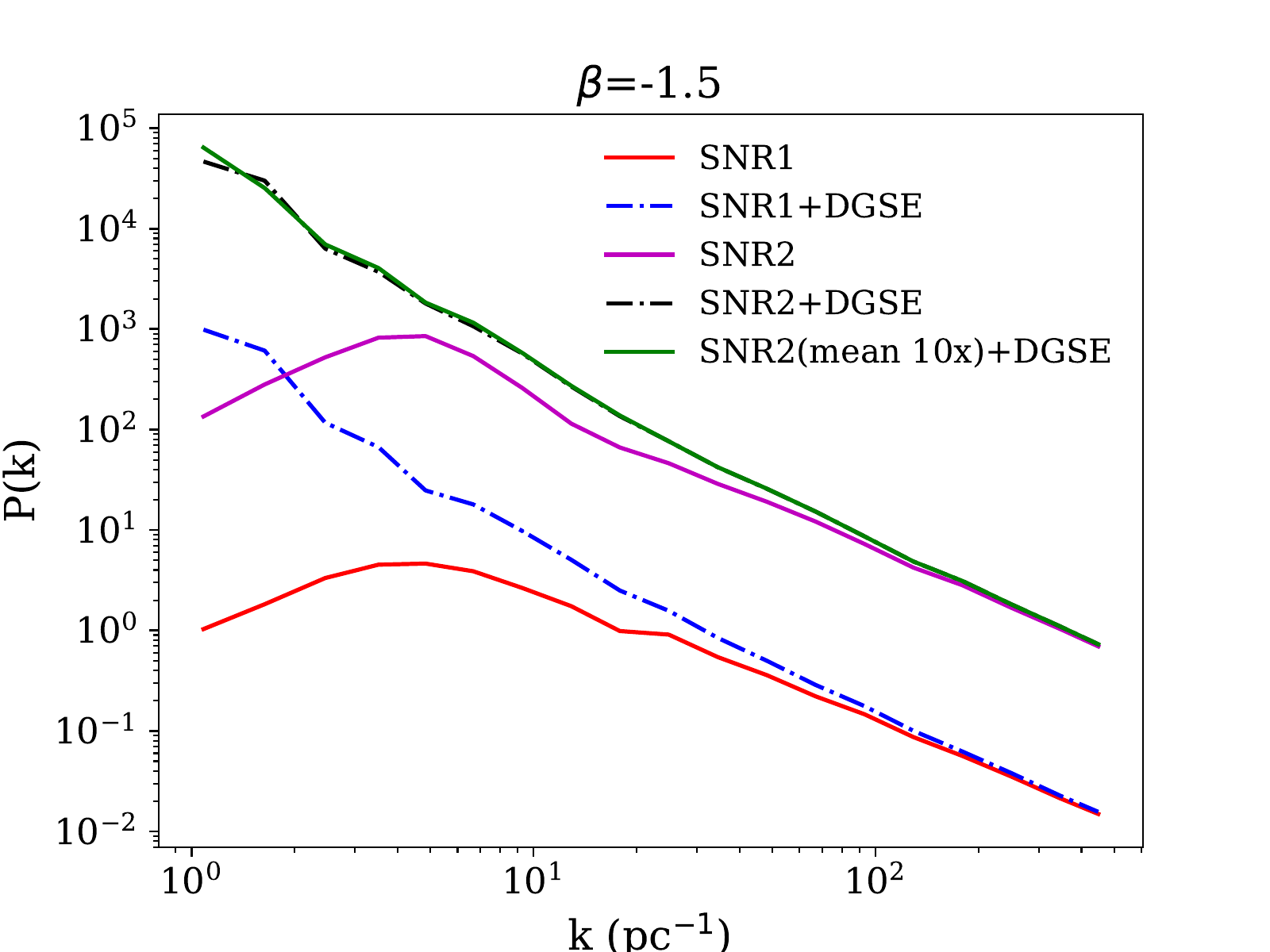}
\includegraphics[width=80mm,angle=0]{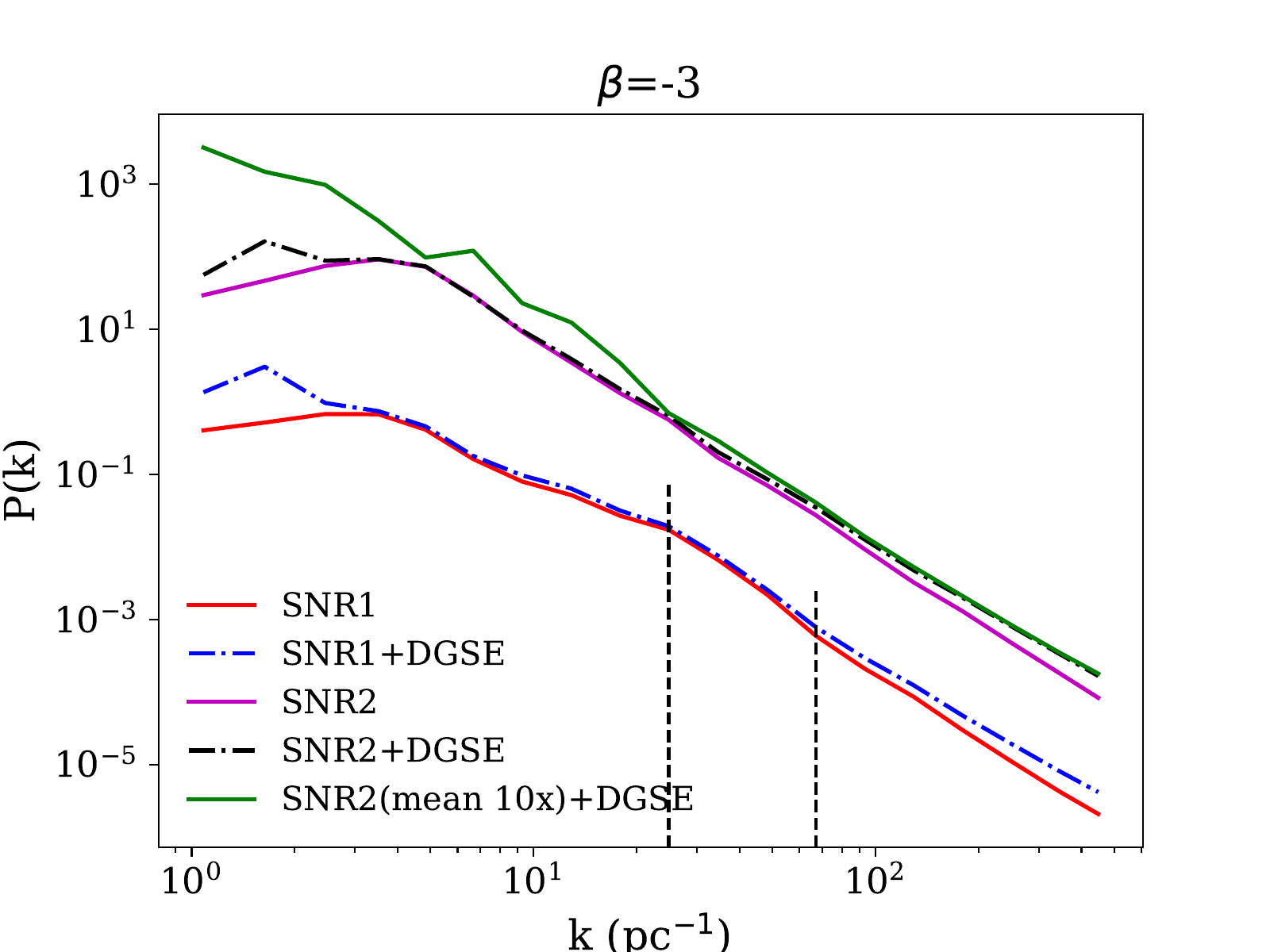}
\caption{The derived SNR power spectra in presence of the DGSE. The left and rights panels are for two different power spectral index of the SNR (-1.5 and -3.0, respectively). We use two different SNR with shell thickness 10 (SNR1) and 30 pixels (SNR2). A DGSE components of power spectral index -2.34 is added with the SNR. The inherent SNR power law power spectrum dominates at large $k$ when the DGSE power spectrum is steeper (left panel). When the DGSE power spectrum is shallower, the input SNR power law power spectrum is recovered only over a small range of $k$, as shown by the two vertical lines in the right panel.}
\label{fig:fig5}
\end{center}
\end{figure*}

The left and right panels of Figure \ref{fig:fig5} summarize the power spectrum results for $\beta=-1.5$ and $-3.0$ respectively. We use SNRs with shell thickness 10 (``SNR1'') and 30 (``SNR2'') pixels, and choose the ballpark DGSE power spectra amplitude such that the contribution is comparable at some $k$ value within the range of our interest. In the left panel of Figure \ref{fig:fig5}, we show the estimated power spectra for the SNR power law (index -1.5), which is shallower than the DGSE power law (index -2.34). The lower curves are SNR-only (red solid line) and SNR+DGSE (blue dashed-dot line) simulations for a shell thickness of 10 pixels (SNR1). Here also, there is a break in the power law, and the lower $k$ range below the break is dominated by the DGSE. However, unlike the break due to the geometrical effect, here, the spectrum is shallower above $k_{\rm break}$. The upper curves are for SNR shell thickness of 30 pixels (SNR2), and the behavior is quite similar. As expected, $k_{\rm break}$ depends on the relative amplitude of the SNR and the DGSE power spectra. The right panel, on the other hand, shows the case when the SNR power law (index -3.0) is steeper than the DGSE power law. The solid red line in the right panel is from the SNR-only simulation for SNR1. As discussed earlier, due to shell thickness and projection effect, the spectrum deviates from the input one at $k\lesssim25~{\rm pc^{-1}}$. However, after including the DGSE, the spectrum (blue dashed-dot line) deviates to become shallower at large $k$ depending on the relative amplitude ($k>70$ for the adopted choice of parameters in Fig~\ref{fig:fig5}). So, only over a small range of $k$ (e.g., $25<k<70~{\rm pc}^{-1}$ in this case, marked by the vertical lines), we recover the original input power law. As shown in the other two lines (SNR2 and SNR2+DGSE), we see a similar behaviour for a larger shell thickness, however a larger $k$ range showing the inherent power-law power spectrum. Recently, \citet{saha19} have observed similar features in the observed intensity fluctuation power spectra of the Kepler SNR at 1.5 GHz (L-band) and 5 GHz (C-band) VLA observations. They found three different regions in the measured power spectrum, as shown in the right panel of Figure \ref{fig:fig5}, with a small $k$ range modified due to geometric effects and large $k$ range having significant contributions from stray foreground emission.

Next we study the effect of the mean surface brightness of the SNR in the measured power spectrum. As mentioned, the power spectrum at lower $k$ values will be modified by this mean component in the LoS collapsed image. We select a particular case from Figure \ref{fig:fig5}; SNR2+DGSE (black dash-dot line in both panels). We increase the mean brightness 10 times as compared with the earlier case (green lines in both panels in Figure \ref{fig:fig5}). For $\beta=-1.5$, when the DGSE is stepper than SNR (left panel), the the lower $k$ values are mostly dominated by the DGSE, so the power spectrum amplitude remain quite similar after increasing the SNR's mean surface brightness. For $\beta=-3$, the lower $k$ values are dominated by the SNR and we see that the power spectrum amplitude increased by an order of magnitude in this $k$ range.

\section{Summary and Conclusions}
\label{discuss}

The study of the power spectrum of the SNR intensity fluctuations at different wavebands can reveal the structures present at sub-pc scales, and also constrain the physical process that generates those small scale structures. Various effects, such as the width of the shell of a shell-type SNR, the projection of the three-dimensional structures in a two-dimensional observational plane, change the observed power spectrum from the intrinsic one. The presence of foreground emission like the Galactic synchrotron emission within which the SNR is embedded can also modify the observed SNR power spectrum. Here, we study these effects and try to outline a method to recover the intrinsic SNR power spectrum by simulating and modeling these effects. We model the SNR as a radially symmetric shell-like structure, with small scale emission fluctuations as three-dimensional GRF with a power-law power spectrum. Although in literature, there exists various detailed hydrodynamical model to simulate the SNR, our simple model is adequate for studying those observational effects in the measured power spectrum. 

Through these simulations, we show the effect of the shell geometry leading to a three-dimensional to two-dimensional transition of the turbulence approximately at scales larger than the SNR's shell thickness. However, due to the effect of projection of a curved structure onto an observed two-dimensional image plane, it is found that we can still measure the three-dimensional intrinsic power spectrum of the SNR at a larger scale than the shell thickness. The effect of shell thickness and projection changes the input power-law power spectrum to a broken power law, where $k_{\rm break}$ depends on the shell thickness, but does not change with the intrinsic power-law index of the power spectrum. In general, all SNRs are embedded in a Galactic environment, and the stray radiation (e.g. at low radio frequency, the DGSE) significantly modify the estimated power spectrum. For the modeling, we also consider two different cases where the DGSE power spectrum is steeper and shallower than the SNR power spectrum. In the presence of DGSE, the range of $k$, over which the original power spectrum is recovered, is more restricted. For a steeper DGSE, the higher $k$ modes are SNR dominated, whereas, for a shallower DGSE, the SNR power spectrum dominates only over an intermediate range of $k$ depending on the relative amplitude. The mean surface brightness of the SNR can increase the power level at lower $k$ values, whereas the DGSE mean surface brightness contributes only to $k = 0$ mode under the assumption of it being approximately constant over the small field of view.

The methodology described in this paper can be used to model different effects to infer the true intensity fluctuation power spectrum in SNRs from the observed power spectrum. Such intensity fluctuations are believed to be generated via fluctuation in physical quantities driven by turbulence in the SNR shell, and hence constraining the true power spectrum can reveal the nature of the turbulence in the SNR's plasma. For simplicity, here we have shown results only for spherically symmetric case. However, complicated ignition geometry and detonation front may lead to non-spherical morphology for SNRs \citep{ferrand20}. The deviation from the spherical shape of the window for the emitting region (e.g. the SNR 1006 has a bilateral shape \citep{fang20}) can be easily included in the simulation. Such deviation of the window is expected to affect the observed power spectrum at small k (large angular scale) range. The other parameters such as the slope and the amplitude of the SNR power spectrum, the relative amplitude of the SNR and DGSE are also scalable to model the data of a specific SNR. Recently, \citet{saha19} studied the synchrotron emission fluctuation power spectrum in the Kepler SNR, and concluded from similar modeling, that the observed spectrum is consistent with Kolmogorov-like turbulence with $\beta=-4.39$. 

The present analysis considered here mainly focus on radio observations, however the methodology is generic and can be applied to data at other wavebands (e.g., X-ray emission). Recently, Saha et al. (2020) (submitted) have used this approach to model the power spectrum of the Cas A SNR from X-ray data. We plan to extend the forward modeling described here for multiwavelength data of a population of Galactic SNRs to learn more about the nature of turbulence in the SNRs.

\section{Acknowledgements}
SC is supported by a research contract between Queen Mary University of London and the University of California at Berkeley.

\section{DATA AVAILABILITY}
The data from this study  will be shared on reasonable request to the corresponding author.

\end{document}